\documentclass[10pt,conference]{IEEEtran}
\IEEEoverridecommandlockouts
\usepackage[style=ieee, mincitenames=1, maxcitenames=2]{biblatex}
\AtBeginBibliography{\small}
\bibliography{ref}
\usepackage{amsmath,amssymb,amsfonts}
\usepackage[inline]{enumitem}
\usepackage{algpseudocode}
\usepackage{graphicx}
\usepackage{textcomp}
\usepackage{multirow}
\usepackage{framed}
\usepackage{verbatim}
\usepackage{xcolor}
\usepackage{threeparttable}
\usepackage{framed}
\usepackage{balance}
\usepackage{amsmath}
\usepackage{longtable}
\usepackage{url}
\usepackage{color}
\usepackage{listings}
\usepackage{graphicx}
\usepackage{booktabs}
\usepackage{amsmath}
\usepackage{amssymb}
\usepackage{pifont}
\usepackage{algorithm}
\usepackage[frozencache,cachedir=.]{minted}
\usepackage{tikz}
\usepackage{float}
\usepackage{balance}
\usepackage{graphicx}
\usepackage{listings}
\usetikzlibrary{fit,backgrounds}
\newcommand{\xmark}{\ding{55}}  
\newcommand{\cmark}{\ding{51}}  

\definecolor{mygreen}{RGB}{34, 139, 34}  
\definecolor{highlightcolor}{RGB}{255,235,205} 

\definecolor{light-gray}{gray}{0.80}

\lstdefinelanguage{Python}{
    keywords={},
    basicstyle=\ttfamily\footnotesize,
    frame=single,
    numbers=left,
    escapeinside={(*@}{@*)},
    moredelim=[is][\color{red}]{@@}{@@},      
    moredelim=[is][\color{mygreen}]{&&}{&&}     
}

\colorlet{shadecolor}{gray!10}
\colorlet{framecolor}{black}
{\endMakeFramed}
\title{Improved Detection and Diagnosis of Faults in Deep Neural Networks Using Hierarchical and Explainable Classification}

\author{
    \IEEEauthorblockN{Sigma Jahan}
    \IEEEauthorblockA{
        Dalhousie University,\\
        sigma.jahan@dal.ca
    }
    \and
    \IEEEauthorblockN{Mehil B Shah}
    \IEEEauthorblockA{
        Dalhousie University,\\
        shahmehil@dal.ca
    }
    \and
    \IEEEauthorblockN{Parvez Mahbub}
    \IEEEauthorblockA{
        Dalhousie University,\\
        parvezmrobin@dal.ca
    }
    \and
    \IEEEauthorblockN{Mohammad Masudur Rahman}
    \IEEEauthorblockA{
        Dalhousie University,\\
        masud.rahman@dal.ca
    }
}

\begin{document}
\maketitle
\begin{abstract}
Deep Neural Networks (DNN) have found numerous applications in various domains, including fraud detection, medical diagnosis, facial recognition, and autonomous driving. However, DNN-based systems often suffer from reliability issues due to their inherent complexity and the stochastic nature of their underlying models. 
Unfortunately, existing techniques to detect faults in DNN programs are either limited by the types of faults (e.g., hyperparameter or layer) they support or the kind of information (e.g., dynamic or static) they use.
As a result, they might fall short of comprehensively detecting and diagnosing the faults. In this paper, we present DEFault (Detect and Explain Fault) -- a novel technique to detect and diagnose faults in DNN programs. It first captures dynamic (i.e., runtime) features during model training and leverages a hierarchical classification approach to detect all major fault categories from the literature. Then, it captures static features (e.g., layer types) from DNN programs and leverages explainable AI methods (e.g., SHAP) to narrow down the root cause of the fault. We train and evaluate DEFault on a large, diverse dataset of $\approx$~14.5K DNN programs and further validate our technique using a benchmark dataset of 52 real-life faulty DNN programs. Our approach achieves $\approx$~94\% recall in detecting real-world faulty DNN programs and $\approx$~63\% recall in diagnosing the root causes of the faults, demonstrating 3.92\%--11.54\% higher performance than that of state-of-the-art techniques. Thus, DEFault has the potential to significantly improve the reliability of DNN programs by effectively detecting and diagnosing the faults.

\end{abstract}

\begin{IEEEkeywords}
Deep Neural Networks, Dynamic Analysis, Model fault, Static Analysis, Training fault
\end{IEEEkeywords}

\section{Introduction}

In recent years, Deep Neural Network (DNN)~\cite{montavon2018methods} applications have been used in many critical domains, including but not limited to fraud detection~\cite{mubalaike2018deep}, software debugging~\cite{mahbub2023, mahbub2024}, medical diagnosis~\cite{esteva2019guide}, facial recognition~\cite{liu2017sphereface}, and autonomous driving~\cite{grigorescu2020survey}. Their widespread adoption marks a major shift towards Software Engineering 2.0~\cite{dilhara2021understanding}. In this new paradigm, intelligent components powered by DNN models are integral to software systems, significantly influencing software development and maintenance practices~\cite{devanbu2020deep}. However, despite the increasing popularity and success, DNN-based applications still suffer from reliability issues, and their faults are harder to detect and resolve compared to traditional software faults~\cite{jahan2024towards}. In traditional software systems, the control flows and data flows can be inspected and analyzed to reason about an underlying fault~\cite{ribeiro2018jaguar}. However, DNN models rely on high-dimensional tensors with millions or billions of parameters, where individual values may lack representation compared to the tensor as a whole~\cite{mahbub2023thesis}. Moreover, DNN models are stochastic and may yield different results across trials~\cite{abdolrasol2021artificial}. All these fundamental differences between traditional and DNN programs render traditional techniques ineffective~\cite{tensorflowprogrambugs} for detecting faults or errors in DNN programs. 

In the context of DNN, training and model faults are most prevalent~\cite{dlbugcharacterstics, faulttaxonomy, jahan2024towards}. Training faults occur during the training process due to several issues such as gradient explosion, vanishing gradient, overfitting, and underfitting~\cite{ben2023testing, faulttaxonomy, morovati2023bugs}. On the other hand, model faults arise from poor model architectures, such as incorrect or missing layer properties and flawed neuron connectivity~\cite{chen2023toward}.

To detect and diagnose faults in DNN programs, recent studies rely on two types of analysis, namely -- static and dynamic analysis.
Static analysis focuses on examining a DNN program's structure and syntax without execution~\cite{debar}, whereas dynamic analysis leverages runtime information captured during model training and inference~\cite{deeplocalize}.
\textcite{Neuralink} propose NeuraLint, a static analysis-based technique that transforms the DNN program into a graph and employs predefined rules to detect faulty patterns. \textcite{debar} propose DEBAR that detects numerical faults from a DNN program based on several static analyses such as tensor abstraction, interval abstraction, and affine relation analysis.

Although static methods might be good for several fault types (e.g., API bugs, model bugs)~\cite{morovati2023bugs, Neuralink}, they alone would fall short in capturing the intricate dynamics and runtime behaviors of DNN programs. Dynamic analysis-based approaches, such as UMLAUT~\cite{umlaut} and AutoTrainer~\cite{autotrainer}, utilize rule-based approaches to detect faults. \textcite{deepdiagnosis} propose DeepDiagnosis, a tree-based technique using predefined rules to guide developers from bug symptoms to likely causes. DeepLocalize~\cite{deeplocalize} traces numerical errors back to the faulty layer using dynamic analysis, such as loss in the validation set. Despite the progress made by current approaches in detecting and diagnosing faults of DNN programs, they still suffer from several limitations, as follows.

\textbf{(a) Lack of comprehensive support for root cause analysis of faults in DNN programs:}
According to \textcite{deepcrime}, faults in DNN programs can be classified into seven major categories (Table \ref{tab:seededFaults}) related to model construction and training. Existing techniques can analyze the root causes of only a subset of these categories of faults in the DNN programs. While some of them focus on model-related faults~\cite{Neuralink}, others focus on training-related faults~\cite{autotrainer, deepfd, cradle, umlaut}, but rarely both. More importantly, none of the existing techniques supports weights or regularization-related faults, which encompass 8.3\% of all DNN faults.
Furthermore, the datasets constructed using the existing mutation frameworks~\cite{deepcrime, deepmutation} might not be comprehensive and can represent only a subset of all possible faults.
Consequently, techniques~\cite{deepfd} based on them could identify only a subset of faults in the DNN programs.
Therefore, the support from existing methods might not be comprehensive enough for root cause analysis of many faults in the DNN programs.

\textbf{(b) Partial use of available information:}
Dynamic analysis could be effective in diagnosing certain training-related faults (e.g., incorrect learning rates and wrong loss functions). However, it alone is not sufficient for pinpointing the structural faults of a DNN model. Recent techniques~\cite{deepfd, autotrainer, deepdiagnosis, deeplocalize, umlaut} rely solely on dynamic analysis, overlooking the potential of static analysis in identifying model-related faults. Static analysis can capture the network architecture, layer configurations, and model parameters before training~\cite{tensorflowprogrambugs}. Thus, it can provide valuable insights for detecting structural faults where the existing techniques might fall short.

In this paper, we present a novel technique -- DEFault (Detect and Explain Faults) -- to detect and diagnose faults in DNN programs. It leverages both static and dynamic properties of DNN programs in a hierarchical classification framework to first detect their faults and then diagnose their root causes. Our solution can address the aforementioned challenges from existing works, which makes our work novel. First, DEFault can detect and explain model-related faults from DNN programs, offering a more comprehensive technique for fault localization. Second, it is trained on a more diverse and larger dataset than that of the state-of-the-art technique~\cite{deepfd}. Unlike existing techniques that rely on either dynamic or static analysis for diagnosing faults, we have combined the best of both. Thus, our approach is better suited to detect and diagnose various types of faults in DNN programs.

We trained DEFault with our constructed dataset, which consists of 14,652 DNN programs (9,855 faulty + 4,797 correct) mutated from real-world Feed-Forward Neural Network (FFNN), Recurrent Neural Network (RNN), and Convolutional Neural Network (CNN). 
Then, we evaluated our technique on a benchmark dataset~\cite{deepfd} containing 52 real-world DNN programs from StackOverflow (SO) and GitHub.
We found that DEFault can identify faulty DNN models with 3.92\% higher accuracy than the state-of-the-art technique~\cite{deepfd}. It can also diagnose both training and model-related faults with 11.54\% higher accuracy. 

We thus make the following contribution in this paper:

\begin{enumerate}[label=(\alph*)]
\item A novel hierarchical, explainable classification technique to detect faulty DNN programs and diagnose their root causes, leveraging static and dynamic analyses. An explainer module to explain the root causes of model-related faults using static features. 

\item A new, diverse dataset containing $\approx$~14.5K DNN programs ($\approx$~9.5K faulty programs) targeting seven categories of faults in FFNN, RNN, and CNN. 

\item Proposed five novel dynamic features to predict faults in DNN programs, significantly improving fault prediction performance by $\approx$~5.6\% in terms of recall.

\item Comprehensive evaluation of the proposed technique and comparison with multiple baselines~\cite{deepfd, deeplocalize, autotrainer, umlaut}.

\item A replication package containing our working prototype and experimental data for the replication~\cite{ICSE2025_repo}.

\end{enumerate}

\section{Motivating Example}
\label{sec: motivation}
To demonstrate the effectiveness of our technique -- DEFault, let us consider the example in Listing~\ref{lst:layerbug}, taken from StackOverflow post \#50079585.
This faulty DNN program attempts to measure the severity of car crashes (minor, moderate, or severe) from input images.
Unfortunately, it achieves an accuracy of $\approx$~32\%.
According to the accepted answer on StackOverflow, the model has three faults. First, there should be exactly three neurons in the final \texttt{Dense} layer as there are three output classes.
Second, the selected activation function \texttt{sigmoid} does not work well when the number of classes is more than two, and it should be replaced with \texttt{softmax} function.
Finally, the selected loss function should be replaced by \texttt{categorical\_crossentropy}, which is more appropriate for multi-class classification. 
\vspace{-0.5em} 
\begin{listing}[h]
\begin{minipage}{\linewidth}
\setminted{
    breaklines,
    fontsize=\scriptsize,
    style=tango,
    highlightcolor=highlightcolor, 
    highlightlines={10,11,13},
    escapeinside=||,
    framesep=2mm, 
    baselinestretch=1.0,
    breakanywhere, 
    breakautoindent=true
}
\begin{minted}{python}
model = Sequential()
model.add(Conv2D(32, (3, 3), input_shape=input_shape))
model.add(Activation('relu'))
model.add(MaxPooling2D(pool_size=(2, 2)))
…
model.add(Flatten())
model.add(Dense(64))
model.add(Activation('relu'))
model.add(Dropout(0.5))
model.add(Dense(1))
model.add(Activation('sigmoid'))

model.compile(loss='binary_crossentropy',
              optimizer='rmsprop',
              metrics=['accuracy'])
\end{minted}
\end{minipage}
\caption{A Faulty DNN Program from StackOverflow}
\label{lst:layerbug}
\end{listing}
\vspace{-1em}

Among the existing techniques for fault detection, AutoTrainer~\cite{autotrainer} fails to detect any issues, probably due to its rule-based design, which only identifies patterns matching predefined symptoms for specific training problems -- none of which were present in the motivating example. UMLAUT~\cite{umlaut} incorrectly diagnoses the fault as `missing activation functions' since its heuristics do not align with the actual program fault. DeepLocalize~\cite{deeplocalize} fails to locate any fault as it is specifically designed to locate numerical errors (e.g., NaN or infinity in loss calculation) that were not present in the example model. Similarly, DeepFD~\cite{deepfd} was not designed to handle layer-related faults and also fails to detect the loss function fault. In contrast,  our technique, DEFault, integrates both static and dynamic analyses, enabling it to comprehensively detect and diagnose faults. Using dynamic features, DEFault detects the incorrect loss function. By leveraging static features with the explainer module, our tool finds the inappropriate activation function (\texttt{sigmoid} instead of \texttt{softmax}) and the incorrect neuron count in the final Dense layer.

\section{Background}\label{sec:background}
In this section, we discuss the most relevant existing approaches and attempt to place our work in the literature. Existing approaches adopt the following methodologies: static analysis, dynamic analysis, mutation-based methods, and adversarial techniques.\\
\indent
\textit{Static analysis-based} techniques detect faults or errors in the DNN programs by analyzing their structures and syntax without execution. Tools like NeuraLint~\cite{Neuralink} use graph transformations and predefined verification rules to identify design inefficiencies like incompatible loss functions or poorly configured layers. Similarly, NerdBug~\cite{nerdbug} adopts an abstraction-based approach to efficiently detect structural and coding bugs (e.g., API misuses and logic errors). DEBAR~\cite{debar} employs tensor abstraction and interval analysis to detect numerical bugs in neural architectures. \\
\indent
\textit{Dynamic analysis-based} techniques capture model behavior during model training, complementing static approaches. For instance, CRADLE~\cite{cradle} detects bugs in DL libraries through cross-backend testing and anomaly tracking. Tensfa~\cite{tensfa} uses decision trees to classify crash messages and performs shape tracking and data dependency analysis to debug tensor shape faults. MODE~\cite{mode} employs state differential analysis to identify faulty neurons causing misclassification, addressing overfitting and underfitting issues. DeepDiagnosis~\cite{deepdiagnosis} maps anomalies in training metrics (e.g., vanishing/exploding gradients) to potential root causes using predefined rules. AutoTrainer~\cite{autotrainer} extends this by detecting and repairing common training issues like exploding gradients and slow convergence in real-time. DeepFD~\cite{deepfd} uses a learning-based approach to classify five common training faults with high accuracy. DeepLocalize~\cite{deeplocalize} locates numerical errors in faulty layers or hyperparameters by analyzing runtime values like weights, gradients, and loss. UMLAUT~\cite{umlaut} provides debugging recommendations by heuristically mapping runtime symptoms to fault categories. GRIST~\cite{GRIST} uses gradient backpropagation to diagnose numerical faults.\\
\indent
\textit{Mutation-based} and \textit{adversarial method-based} techniques detect faults in the DNN models by introducing controlled perturbations or modifications to models. DeepMuFL~\cite{deepmufl} employs mutation analysis and detects faulty mutants by ranking them based on their fault suspiciousness scores. Adversarial approaches like DeepFault~\cite{deepfault} introduce perturbations to highlight model vulnerabilities, such as neuron-specific robustness issues.\\
\indent
Our proposed technique, DEFault, uses a \textit{hybrid learning-based} approach that integrates static and dynamic analyses within a hierarchical classification framework, enabling comprehensive fault detection and diagnosis. DEFault also supports a wider range of faults (seven main categories out of eight from the DL fault categories), including those arising during model construction and training. Furthermore, by incorporating SHAP-based explanations, DEFault enhances interpretability, enabling precise diagnosis of faults in both structural and training contexts.

\section{DEFault}
\subsection{Collection of Faulty DNN Programs}
Collecting a large number of correct and faulty DNN programs to train a fault detector is a major challenge. Thus, we developed a systematic approach to collect DNN programs from StackOverflow, inspired by prior works~\cite{deepfd, deep4deep}. We leverage the tagging system and content-based criteria of SO. The criteria include the recency of a question (last three years), the presence of accepted answers (at least 1), and the answer being of high quality (5 or higher upvotes)~\cite{shah2024towards}. We limited our search to \textit{TensorFlow} and \textit{Keras} to align with popularity and recent literature~\cite{deepfd, deepdiagnosis, deeplocalize}. After this step, we got 328 DNN programs. 
We then manually reviewed each DNN program and corresponding SO post containing a model structure (e.g., improper layer configurations) or training (e.g., over-fitting) related fault. 
For example, the errors concerning import statements or API types might not be directly related to the training or model aspects of the DNN programs. These errors often stem from the environment setup, package versioning, or API usage rather than training steps or model structures. We also carefully examine the SO posts to verify that they contain sufficient information about faulty DNN models, including symptoms, faults, code snippets, and potential solutions. After all the filtration steps above, we got 89 DNN programs. 

\begin{figure*}[h]
  \centering
  \includegraphics[width=0.90\textwidth]{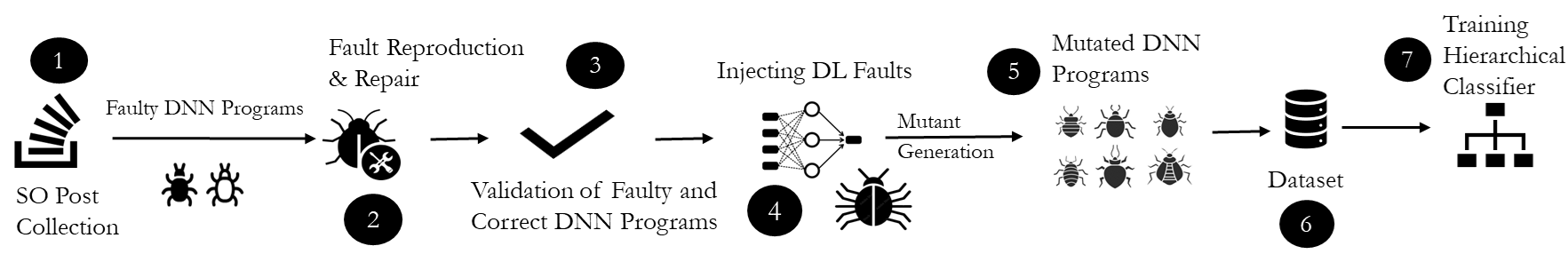}
  \vspace{-1em}
  \caption{Schematic diagram of dataset preparation and classifier model training}
   \vspace{-2em}
  \label{fig:dataprep}
\end{figure*}

\subsection{Reproduction and Repair of Faulty DNN Programs} 
After collecting 89 faulty DNN programs, we conducted another extensive manual analysis where we reproduced and repaired them. First, we determine the feasibility of reproducing each bug by executing the corresponding DNN programs. If the DNN program crashed due to API upgrades, versioning, or typo issues, we fixed them. Then, we execute the code and attempt to reproduce the symptoms of bugs reported at SO. After this step, we found 60 DNN programs reproducing the bugs, which were used for subsequent analysis. For each of the 60 reproducible faulty programs, we carefully studied the corresponding SO post, including the problem description, code snippets, and the accepted answer, to understand the root cause of the bug and the suggested fix. We then applied the fix to the faulty model and re-ran the code to verify whether the bug was successfully repaired. In cases where the suggested fix from the SO post did not completely resolve the issue, we further investigated the problem and performed additional modifications until the faulty program was fully repaired. After repairing each faulty model, we designed specific test cases based on the expected behavior described in the SO post. We then tested the fixed version with the provided dataset or a suitable alternative (e.g., dummy dataset). We compared the output of the repaired model with the expected results to confirm that the bug was resolved and that the model was functioning as intended. We successfully repaired all 60 faulty DNN programs in our dataset.

\subsection{Validating Faulty and Correct Models}
To validate the manual selection, filtration, reproduction, and repair of faulty DNN programs, two authors of this work independently reviewed the SO posts at each stage of the process. During the initial filtration stage, both authors examined all potential faults to determine their eligibility for inclusion in the study. They assessed whether each post met the criteria for advancement to subsequent stages. After completing their assessments, the authors discussed their findings to resolve any discrepancies and reach a consensus with 83.2\% value of Cohen Kappa, indicating a strong level of agreement~\cite{sun2011meta}. For the reproduction and repair phases, the first author and the second author both conducted the process, starting with 5 bugs from the dataset. After achieving an initial Cohen Kappa of 61.5\%, they held two meetings to identify and resolve the main reasons for their disagreements. In the next round, they worked with 10 more bugs and achieved a Cohen Kappa of 82.8\%, which is considered an almost perfect agreement~\cite{sun2011meta}. Subsequently, the first author reproduced the remaining 45 bugs while the second author checked the reproduction, achieving an average Cohen Kappa of 85.4\%.

\begin{table*}[t]
\centering
\caption{Fault Injection Supported by DEFault and DeepCrime}
\label{tab:seededFaults}
\vspace{-1em}
\begin{tabular}{|c|ccccccc|ccc|}
\hline
\begin{tabular}[c]{@{}c@{}}Fault Group\end{tabular} & Hyperparameter & Layer & Loss & Activation & Optimization & Weights & Regularization & FNN & CNN & RNN \\ \hline \hline
DEFault* & \ding{51} & \ding{51} & \ding{51} & \ding{51} & \ding{51} & \ding{51} & \ding{51} & \ding{51} & \ding{51} & \ding{51} \\ \hline
DeepCrime  & \ding{51} & \ding{55} & \ding{51} & \ding{51} & \ding{51} & \ding{51} & \ding{51} & \ding{51} & \ding{51} & \ding{55} \\ \hline
\end{tabular}
\vspace{-1em}
\end{table*}

\subsection{Injecting Faults into DNN Programs} 
Due to the scarcity of real-world faulty DNN programs, we use mutation-based techniques to generate synthetic data~\cite {deepcrime}. This approach allows us to systematically introduce a wide variety of faults into the DNN programs. There have been a few frameworks to inject faults in DNN programs. DeepMutation~\cite{deepmutation} can inject several types of faults into the layers and weights of the DNN program.
However, these fault types may not be representative of the diverse range of faults encountered across different architectures (e.g., FFNN, RNN, CNN). DeepCrime~\cite{deepcrime} is the state-of-the-art technique for fault injection that can inject six categories of faults in DNN programs. However, it cannot inject any fault into the layers of DNN models. Our manual analysis shows that 21.67\% of DNN faults could be associated with the layer.
Thus, DeepCrime might not be sufficient to generate mutant models containing layer-related faults.
Moreover, it does not support the creation of mutants from RNNs. To address these limitations, we extended DeepCrime in two regards. First, we added ten mutation operators to inject faults in the layers of DNN models, following the guidelines from the modern taxonomy of faults in DNN programs~\cite{faulttaxonomy, dlbugcharacterstics, deepcrime, tensorflowprogrambugs}. We apply these mutation operators by systematically selecting appropriate layers. For instance, to apply the \textit{change activation function} operator, we randomly choose from the layers with an activation function and replace it with another function, randomly chosen from a set of 11 standard activation functions (e.g., ReLU, Sigmoid). Similarly, we apply regularization and architectural mutations directly to compatible layers to ensure the diversity of the mutated models. We carefully limit all mutation parameters within practical boundaries (e.g., CNN kernel sizes between 1–7) in accordance with the literature~\cite{Goodfellow-et-al-2016, Johnson-2017} to maintain realistic and functional model configurations. Details can be found in our replication package~\cite{ICSE2025_repo}.
\begin{figure}[h]
\vspace{-1.0em}
\centering
\includegraphics[width=0.85\columnwidth]{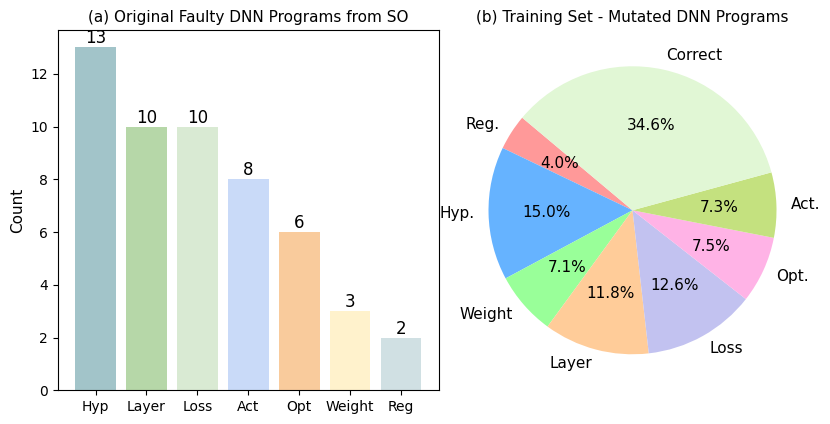}
    \vspace{-1em}
    \caption{Distribution of fault types in our datasets}
    \vspace{-1em}
    \label{fig:faultdist}
\end{figure}

This extension ensures comprehensive coverage of fault types in DNN programs. Then, we also add support for RNNs with all fault types. We use \texttt{isKilled()} from the DeepCrime method to determine if a mutated model is faulty. It works by comparing the mutant's accuracy distribution with that of the original model on a testing set. We use statistical tests (Generalized Linear Model~\cite{ialongo2016understanding}) to assess the significance and effect size (Cohen's d~\cite{lee2016alternatives}) of the difference between the two accuracy distributions. If the mutant's accuracy distribution is significantly different and worse than the original model's, it is considered faulty and labeled accordingly; otherwise, it is considered correct.
\begin{figure*}
    \centering    
    \includegraphics[width=0.80\textwidth]{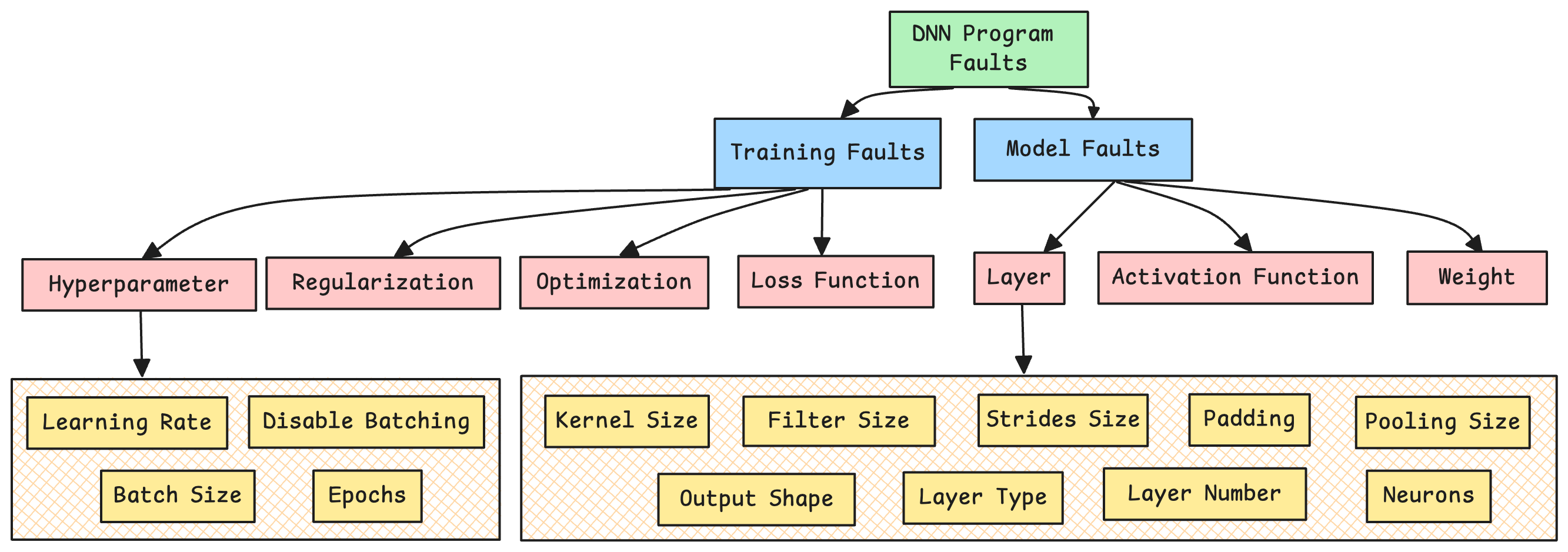}
    \vspace{-1em}
    \caption{Types of DNN faults and their root causes}
    \label{fig:DNNFaultType}
    \vspace{-1.5em}
\end{figure*}

Inspired by mutation testing research~\cite{deepcrime, deepmutation, deepmutation++}, we use the mutation score to check their applicability to different architectures (e.g., FFNN, CNN, RNN). The mutation score is defined as the proportion of mutation operator instances killed by the training set and the test set over all those killed by the training set. A high mutation score indicates that the new mutation operators effectively create meaningful faults that both the training and test sets can detect. The experiment shows high mutation scores (0.85 for FFNN, 0.83 for RNN, and 0.87 for CNN) for our mutant models. In real world, a single DNN program might have multiple faults. To create mutants with multiple faults, we adopt heuristics inspired by DeepFD~\cite{deepfd} and inject an average of three and a maximum of seven types of faults in a single mutant (i.e., DEFault supports seven categories of faults). The process involves sequentially injecting faults one after another. We then validated the injected faults using the same technique (i.e., \texttt{isKilled()} method~\cite{zhu2021kill}). This approach ensures that all mutations remain mathematically valid, operationally sound, and representative of real-world scenarios.

Table \ref{tab:seededFaults} presents the fault categories used for injecting faults into DNN programs.
Due to the stochastic nature of DNN programs, we follow prior works~\cite{deep4deep, deepfd, autotrainer} to reduce the randomness and retrain and evaluate each model 15 times.
We focused on seven key fault categories from DeepCrime~\cite{deepcrime} (see Fig. \ref{fig:DNNFaultType}), excluding data-centric mutation operators, as our work focuses on the faults encountered during model construction and training rather than from training data quality. Note that categories like Weights and Regularization affect DNN's training process globally~\cite{Skabar2005, Advani2020}, making their further diagnosis redundant. In contrast, Layers and Hyperparameters have localized effects, making their fine-grained subcategories (e.g., filter size, learning rate) essential. Therefore, we further divide only Layer and Hyperparameter faults into their subcategories for effective root cause analysis.

\subsection{Preparation of Training Dataset}
We collect 60 DNN programs, apply the mutation operators (refer to Table \ref{tab:seededFaults}), and capture 14,652 mutants. We also apply killability criteria, and collect 9,855 faulty models and 4,797 correct models. During mutant creation, we capture their static and dynamic features as follows. 

\subsubsection{Static Feature Extraction} 
\label{ssse: static-feature}
Static features are characteristics of a DNN model that can be extracted by analyzing the source code without executing the code~\cite{hu2020deepsniffer}. These features provide insights into the structure, complexity, and design choices of the DNN model~\cite{Neuralink}. We collect five categories of static features as shown in Table~\ref{tab:static_features}. All these features, proposed by the existing literature~\cite{Neuralink, shen2021comprehensive, dlbugcharacterstics}, showed their potential to identify faults in DNN programs. Furthermore, there exists a direct mapping between static features to specific DNN code sections, which could be faulty~\cite{zhang2021understanding, nguyen2019understanding}. 
As a result, these static features could effectively diagnose the root cause of a fault in the DNN model (see Section~\ref{subsec: explainable framework}). 
\begin{table}[h]
\centering
\caption{Static Features Extracted from DNN Models}
\vspace{-1em}
\label{tab:static_features}
\resizebox{\columnwidth}{!}{
\begin{tabular}{|l|p{5.2cm}|}
\hline
\textbf{Feature Components} & \textbf{Description} \\
\hline
\hline
Layer Counts & Diversity. Count of different layer types and the unique type of layers (e.g., Dense, Conv2D, LSTM) \\
\hline
Neurons Count & Number of neurons in all applicable layers \\
\hline
Input Shapes & Dimension of the input layers \\
\hline
Output Shapes & Dimension of the output layers \\
\hline
Mismatch Detection & Detecting mismatches in input and output dimensions between consecutive layers \\
\hline
\end{tabular}}
\vspace{-2em}
\end{table}

\begin{figure*}[t]
  \centering
  \includegraphics[width=\textwidth]{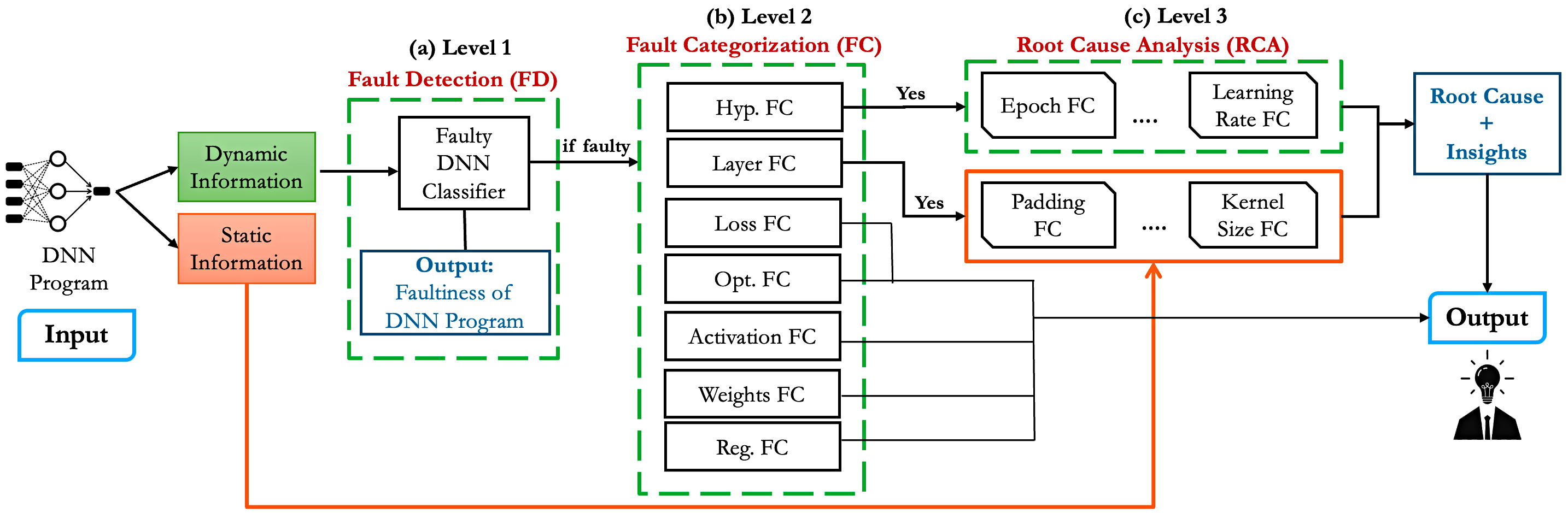}
    \vspace{-2em}
   \caption{Workflow of DEFault: (a) Fault detection, (b) Fault categorization, and (c) Root cause analysis}
  \label{fig:schematic_diagram}
  \vspace{-1.5em}
\end{figure*}

\subsubsection{Dynamic Feature Extraction} 
\label{ssse: dynamic-feature}
Dynamic features, which are extracted during model training, can provide valuable insights into the symptoms and root causes of faulty DNN models~\cite{deepfd, deeplocalize, deepdiagnosis, cradle, autotrainer}.
DEFault collects 23 runtime features (Table~\ref{tab:runtimedata}), periodically captured at every epoch, using our custom callback methods implemented within each DNN program. Among them, 17 features were taken from existing literatures~\cite{deepfd, deeplocalize, autotrainer} and are proven effective in detecting faults in DNN programs. The remaining six are our novel features. The choice of these novel features is grounded in practical observations and validated by research in deep learning~\cite{bengio2012practical, smith2017cyclical, pinto2009high}. 

\textit{Activation:} Activation saturation can limit a model's learning capacity~\cite{deepdiagnosis}. To calculate the percentage of saturated activation values at each epoch, we collect activation statistics and compare them against predefined thresholds, determining if they are saturated at their minimum or maximum value. This helps identify potential learning bottlenecks caused by activation saturation.

\textit{Learning Rate:} DEFault tracks the adjusted learning rate at each epoch when the learning rate is scheduled or adaptive. This feature helps categorize hyperparameter and optimizer faults that might prevent model convergence~\cite{smith2017cyclical}. We collect the adjusted learning rate value at each epoch, considering any scheduling or adaptations applied during training.

\textit{Hardware Utilization:} We track the hardware utilization metrics, including CPU, GPU, and memory usage. These hardware-specific metrics help us identify potential faults and inefficiencies~\cite{pinto2009high}. Abnormal hardware utilization metrics can indicate inefficient memory management, including too much communication between CPU and GPU or unnecessary type conversions, leading to computational bottlenecks or even higher numerical error~\cite{glorot2010understanding}. We measure these metrics regularly during training using system monitoring tools or APIs provided by the DNN framework (e.g., TensorFlow, Keras). 

\subsection{Training of Multi-Level Classifier}
Once the training dataset is prepared, we train a hierarchical classification approach~\cite{arabie1996hierarchical} on the dataset to detect and categorize faults in DNN programs. We choose the Random Forest~\cite{rigatti2017random} classifier as the base model due to its ability to handle high-dimensional data and capture non-linear relationships~\cite{breiman2001random}.
Our approach consists of multiple levels of classifiers.
First, we identify whether a DNN program contains faults or not (see Section~\ref{sssec: fault-detection}).
If yes, a set of binary classifiers identifies the specific fault category among the seven groups: Hyperparameter, Loss, Activation, Layer, Optimizer, Weight, and Regularization (see Section~\ref{sssec: fault-cat}). Since a single DNN program can have multiple faults (e.g., see Section~\ref{sec: motivation}), we use seven binary classifiers to detect one or more fault categories. We do not use multilabel classifiers to avoid severe data imbalance problems. 
Among these categories, hyperparameter faults can emerge from different root causes. Thus, we use four binary classifiers to diagnose the root causes of hyperparameter faults (learning rate, disable batching, batch size, and epochs) at a finer-grained level.

We follow a similar training pipeline for all levels of classifiers. We load and preprocess the data (e.g., scaling, removing null values, encoding categorical values), split it into training, validation, and testing sets, initialize Random Forest classifiers, and train them on the preprocessed data. We also perform a grid search with 5-fold cross-validation to mitigate randomness and to find each classifier's best hyperparameters (e.g., max depth, max features)~\cite{anggoro2021performance}. Grid search optimizes the model's performance by finding the best hyperparameters~\cite{bergstra2012random}. The dynamic features extracted from the DNN programs serve as input to these classifiers.

\subsection{Construction of Explainer Module} \label{subsec: explainable framework}
Once fault categories are detected, we design an explainer module to diagnose their root causes. Since the developers act on the DNN model's structures (e.g., the number of neurons in a layer), we analyze the root cause of layer faults using static features. First, we train a new Random Forest model on the static features to detect whether a DNN program has a layer fault. Then, we use SHAP~\cite{angelov2021explainable} to identify which static properties contributed the most towards the faulty verdict. SHAP can rank the training features based on their importance in a prediction~\cite{mangalathu2020failure}. It determines which static feature is the most likely to cause the specific layer fault.

\begin{table*}[ht]
\caption{Dynamic Features Extracted From DNN model training}
\vspace{-0.5em}
\label{tab:runtimedata}
\centering
\begin{tabular}{|p{3cm}|p{6cm}|p{6cm}|}
\hline
\multicolumn{1}{|c|}{\textbf{Runtime data}} & \multicolumn{1}{c|}{\textbf{Monitored variables}} & \multicolumn{1}{c|}{\textbf{Description}}                          \\ \hline
\hline
Training Metrics & \raggedright loss, val\_loss, train\_acc, val\_acc &
Loss \& accuracy on the training \& validation sets \\ \hline

Weight & \raggedright large\_weight\_count, cons\_mean\_weight\_count, cons\_std\_weight\_count, nan\_weight\_count &
Statistics related to model weights, including large weights, constant mean/std, \& NaN weights \\ \hline

Accuracy and Loss Trends & \raggedright acc\_gap\_too\_big, loss\_oscillation, decrease\_acc\_count, increase\_loss\_count &
Trends in accuracy and loss across epochs, including gaps, oscillations, and changes from previous epochs \\ \hline

Activation & \raggedright dying\_relu, activation\_statistics*, saturated\_activation* &
Activation statistics (dying ReLUs, mean/std, saturated sigmoid/tanh) \\ \hline

Gradient & \raggedright gradient\_vanish, gradient\_explode, nan\_gradients\_count, gradient\_statistics &
Gradient statistics (vanishing/exploding, NaNs, value distribution) \\ \hline
   
Learning Rate & \raggedright adj\_lr* &
Adjusted learning rate for the epoch \\ \hline
Hardware Utilization & 
\raggedright cpu\_utilization*, gpu\_memory\_utilization*, memory\_usage* &
Hardware resource utilization: CPU, GPU memory, and overall memory usage. \\ \hline
\multicolumn{3}{l}{\footnotesize *Proposed by DEFault}
\end{tabular}
\vspace{-2em}
\end{table*}

\subsection{Model Inference} 
\label{subsec:inference}
Fig.~\ref{fig:schematic_diagram} illustrates how DEFault identifies faulty DNN programs and analyzes their root causes. It uses our trained multi-level classifiers for inference. We use the DNN program from Listing~\ref{lst:layerbug} to describe DEFault's inference process.

\subsubsection{Fault detection}
\label{sssec: fault-detection}
At Level 1, we attempt to find out whether a given DNN program is faulty or not. First, we extract the dynamic information from the program, as described in Section~\ref{ssse: dynamic-feature}.
Using the information, the Level 1 classifier predicts whether or not the DNN program has a fault. If the DNN program is faulty, we proceed to categorize the fault.

\subsubsection{Fault categorization}
\label{sssec: fault-cat}
At Level 2, we attempt to categorize the faults in the DNN program. We employ seven different classifiers that leverage dynamic features and target seven fault categories (Fig.~\ref{fig:schematic_diagram}). A DNN program can have one or more categories of faults. Each classifier predicts whether the DNN program has a fault from the corresponding category or not. For hyperparameter and layer categories, we continue with the root cause analysis as they are quite broad categories (check Fig. \ref{fig:DNNFaultType}). For any other category, we stop the analysis.

\subsubsection{Root Cause Analysis}
\label{ssec: root-cause}
We take two slightly different approaches to analyze the root cause of hyperparameter and layer faults.
Hyperparameters of a DNN program are connected to the training phase and thus affect the training behaviors of the model.
Therefore, we employ another set of binary classifiers leveraging the dynamic information to classify hyperparameter faults into more granular levels. These models identify whether the DNN program has faults in learning rate, disabled batching, batch size, or number of epochs. 

On the other hand, the layers of a DNN program are often explicitly defined. 
Therefore, we employ an explainer module (SHAP)~\cite{angelov2021explainable} to identify the static features (see Section~\ref{ssse: static-feature}), potentially explaining the fault.
In the case of the DNN program in Listing~\ref{lst:layerbug}, we find that two static features from our explainer module, representing the number of \texttt{softmax} functions, contribute the most to mark the DNN program as faulty with layer fault.
Since the faulty DNN program does not have any \texttt{softmax} activation function (i.e., count of softmax is 0), this implies that the model needs to have a softmax function.
This way, we identify the root cause of a fault in a DNN program by following a series of inferences.

\section{Experiment}
We curate a large dataset of 14,652 mutated DNN programs and evaluate our technique DEFault -- with standard performance metrics (e.g., accuracy, recall, precision, F1-score). To place our work in the literature, we compare our technique with four relevant baselines~\cite{umlaut, autotrainer, deeplocalize, deepfd}. We also assess the performance of DEFault in detecting and diagnosing 52 real-world faults in DNN programs. In our experiments, we thus answer four research questions as follows.\\
\indent \textbf{RQ$\mathbf{_1}$:} How does DEFault perform in detecting faulty DNN programs and categorizing their fault types?  \\
\indent \textbf{RQ$\mathbf{_2}$:} How does the choice of features affect the performance of DEFault?  \\
\indent \textbf{RQ$\mathbf{_3}$:} Can DEFault outperform the existing state-of-the-art techniques in DNN fault detection \& categorization?  \\
\indent \textbf{RQ$\mathbf{_4}$:} How does DEFault perform in diagnosing the root cause of DNN faults?

\subsection{Experimental Dataset}
Our training dataset consists of 14,652 mutated DNN programs (9,855 faulty \& 4,797 correct). During the model training phase, we split our dataset as 70\%-15\%-15\% for training, validation, and testing. Additionally, to evaluate the effectiveness of DEFault with real-world faults, we used a benchmark dataset~\cite{deepfd} consisting of 52 real-world faulty models obtained from StackOverflow and GitHub.

\subsection{Evaluation Criteria}
\label{subsec: eval-crit}
To automatically evaluate our technique on the test data, we used four widely used performance metrics from ML classification, namely -- accuracy, recall, precision, and F1-score~\cite{wardhani2019cross}.
In our experiment, we define true positive instances following existing studies~\cite{deepfd, umlaut, autotrainer}. A true positive instance for fault detection suggests that detecting the presence of a fault in the DNN program is accurate, regardless of its fault category. A true positive instance for fault categorization suggests identifying the category of the fault correctly (e.g., fault in hyperparameter), regardless of the root cause. For multiple fault types, if at least one fault type is accurately identified of a DNN program, it is considered a true positive~\cite{deepfd}. Finally, a true positive instance for root cause analysis suggests identifying at least one of the root causes of the fault. 

\subsection{Replication of Baseline Techniques}

In our research, we compared our proposed technique -- DEFault -- with four state-of-the-art baselines: UMLAUT~\cite{umlaut}, AutoTrainer~\cite{autotrainer}, DeepLocalize~\cite{deeplocalize}, and DeepFD~\cite{deepfd}. A detailed description of these methods is provided in Section~\ref{sec:background}. To ensure a fair comparison, we directly used their official replication packages~\cite{umlaut, deepfd, deeplocalize, autotrainer} for the replication.  

\subsection{Evaluation of DEFault}
\subsubsection{\textbf{Answering RQ$\mathbf{_1}$} -- Performance of DEFault}
In this experiment, we evaluate the performance of DEFault using appropriate performance metrics in both fault detection and fault categorization. Table~\ref{tab:classification_report} shows the performance of DEFault in detecting faulty DNN programs. 

\setlength{\tabcolsep}{5pt} 
\renewcommand{\arraystretch}{0.9}
\begin{table}[h]
\centering
\vspace{-1em}
\caption{DEFault's Performance on Fault Detection}
\vspace{-0.5em}
\label{tab:classification_report}
\begin{tabular}{lcccc}
\toprule
\textbf{Class} & \textbf{Precision} & \textbf{Recall} & \textbf{F1-Score} & \textbf{Support} \\
\midrule
\midrule
Faulty & 0.97 & 0.98 & 0.98 & 1556 \\
Correct & 0.97 & 0.95 & 0.96 & 927 \\
Macro Avg. & 0.97 & 0.97 & 0.97 & 2483 \\
\bottomrule
\end{tabular}
\vspace{-1em}
\end{table}

Table~\ref{tab:model_performance} presents DEFault's performance in categorizing each type of fault in DNN programs. Loss, Optimization, and Activation faults achieved recalls of 0.96, 0.95, and 0.91, respectively. The high performance is attributed to the presence of dynamic features directly relevant to each of these fault types. Using the feature importance metric, we found that for Loss-related faults, dynamic features such as training loss and validation loss, loss oscillation, and increased loss count provide clear indicators. Similarly, Optimization faults benefit from features like gradient vanishing, exploding, NaN gradients, and adjusted learning rate~\cite{zheng2019layer}. Activation faults can be effectively diagnosed using several dynamic features such as dying ReLU, saturated activation, and activation statistics~\cite{wang2019reltanh}. The direct relevance of these fault-specific features enables DEFault to accurately identify issues of these three categories. 

\begin{table}[h]
\centering
\scriptsize
\vspace{-1em}
\caption{DEFault's Performance on Fault Categorization}
\vspace{-1em}
\label{tab:model_performance}
\setlength{\tabcolsep}{5pt} 
\renewcommand{\arraystretch}{0.95} 
\begin{tabular}{l r c c c c}
\toprule
\textbf{Category}      & \textbf{Testset} & \textbf{Accuracy} & \textbf{Precision} & \textbf{Recall} & \textbf{F1-Score} \\
\midrule
\midrule
Loss                   & 555               & 0.97              & 0.97               & 0.96            & 0.96 \\
Optimization           & 330               & 0.94              & 0.93               & 0.95            & 0.94 \\
Activation             & 322               & 0.91              & 0.92               & 0.91            & 0.91 \\
Hyperparameter         & 658               & 0.93              & 0.93               & 0.92            & 0.92 \\
Layer                  & 520               & 0.86              & 0.89               & 0.84            & 0.86 \\
Regularization         & 176               & 0.95              & 0.94               & 0.95            & 0.94 \\
Weights                & 312               & 0.90              & 0.90               & 0.89            & 0.89 \\
\midrule
\textbf{Overall}       &                   & \textbf{0.92}     & \textbf{0.93}      & \textbf{0.92}   & \textbf{0.92} \\
\bottomrule
\multicolumn{6}{l}{\scriptsize Precision, Recall and F1-Score are weighted averages}
\end{tabular}
\vspace{-1em}
\end{table}

In the Hyperparameter and Regularization fault categories, DEFault obtains a recall of 0.92 \& 0.95. Based on feature importance, we found that dynamic features like GPU \& CPU utilization, training metrics (loss, accuracy), gradient statistics, and adjusted learning rate are the most relevant features, which help to detect these training-related fault categories. For Layer and Weights faults, DEFault achieves a recall of 0.84 \& 0.89, slightly lower than other fault types. Diagnosing these faults might be more challenging due to their model-related nature, the complexity of layer interactions, silent failures, non-linear effects, multiple problematic layers, and dependencies on initialization~\cite{dlbugcharacterstics, faulttaxonomy, Neuralink}. Moreover, DEFault achieves high precision values ranging from 0.89 to 0.97 across all fault categories, indicating that when DEFault identifies a fault and categorizes it, it is highly likely to be accurate in its assessment. The results are consistent across both test and validation sets, demonstrating the robustness of DEFault's fault detection and categorization capabilities.

Overall, DEFault demonstrates strong performance in fault detection and categorization across all fault types. As shown in Table~\ref{tab:model_performance}, DEFault achieves an overall accuracy of 92\%, precision of 93\%, recall of 92\%, and F1-score of 92\%. This means that DEFault accurately detected and correctly categorized 92\% of the faults present in the dataset. The high precision indicates that when DEFault identifies a fault, it is highly likely to be correct, while the high recall shows its effectiveness in capturing most of the actual faults. These results highlight DEFault's robust overall capability in automating the fault diagnosis process in DNN programs.

\setlength{\tabcolsep}{5pt} 
\renewcommand{\arraystretch}{0.95} 
\begin{table}[ht]
\vspace{-1em}
\centering
\scriptsize 
\caption{DEFault's Performance in Root Cause Analysis of Hyperparameter Faults}
\vspace{-1em}
\label{table:performancethirdlevel}
\begin{tabular}{p{1.9cm}p{0.8cm}p{0.8cm}p{0.8cm}p{0.8cm}p{1.1cm}}
\toprule
\textbf{Root Causes} & \textbf{Testset} & \textbf{Accuracy} & \textbf{Precision} & \textbf{Recall} & \textbf{F1-Score} \\
\midrule
\midrule
Batch Size       & 210  & 0.96 & 0.97 & 0.92 & 0.94 \\
Learning Rate    & 195  & 0.97 & 0.96 & 0.96 & 0.96 \\
Disable Batching & 118  & 0.92 & 0.95 & 0.89 & 0.92 \\
Epoch            & 135  & 0.94 & 0.93 & 0.92 & 0.92 \\
\bottomrule
\end{tabular}
\vspace{-1em}
\end{table}

\subsubsection{\textbf{Answering RQ$\mathbf{_2}$} -- Impact of novel and existing features on DEFault's performance}

\begin{figure}[ht]
\vspace{-1em}
\centering
\includegraphics[width=0.7\columnwidth]{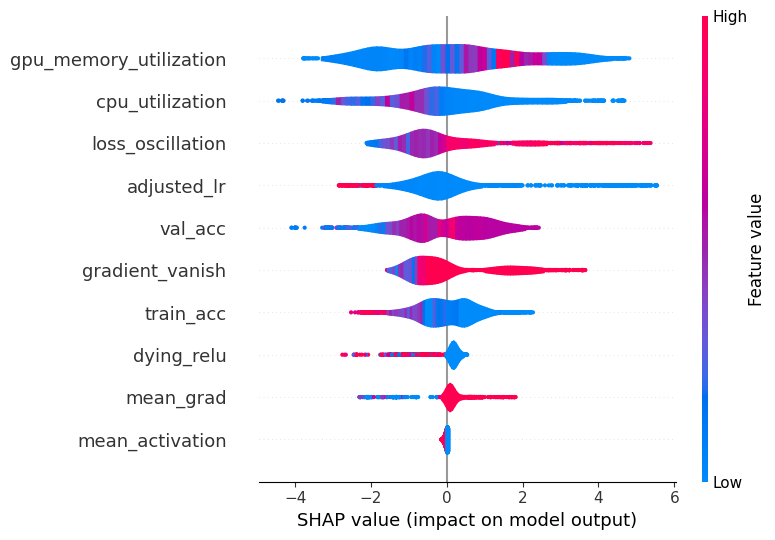}
		\vspace{-1em}
	\caption{Feature Impact on Faulty DNN Program Detection using SHAP}
		\vspace{-1em}
	\label{fig:featureimp}
\end{figure}

We extracted five novel dynamic features from the runtime of DNN programs. To determine the impact of these factors on fault detection, we constructed a Generalized Linear Model~\cite{dunteman2005introduction}. It allows us to test the statistical significance of multiple independent factors in our technique. Moreover, using the global explanation of SHAP, where the absolute Shapley values of all instances in the dataset are averaged~\cite{lundberg2020local}, we found the top five impactful features that drive the model's decisions in determining faulty or correct DNN programs. From Fig.~\ref{fig:featureimp}, we notice a significant contribution of our newly added features, particularly GPU Memory Utilization and Adjusted Learning Rate. GPU Memory Utilization captures unusual tensor operations and data transfer patterns~\cite{liu2017unified}, indicating potential inefficiencies or errors, while Adjusted Learning Rate reflects dynamic changes during training, which are critical for identifying unstable training processes and optimization issues~\cite{wilson2003general}. To further assess the impact of these new features, we conducted an ablation study by removing them from the feature set of our Level 1 classifier (i.e., fault detection). The fault detection performance decreased from 97\% to 93\% (accuracy), confirming the contribution of these five dynamic features to the overall performance of DEFault.

In our approach, we extracted two types of features: dynamic features from the training step (Table \ref{tab:runtimedata}) and static features from the source code (Table \ref{tab:static_features}). To determine which feature type is more effective for fault detection and categorization, we trained our classification models on each type individually. We found that static features alone are not particularly helpful in detecting faulty DNN programs. Using static features only, DEFault achieves 62.52\% accuracy on the test set, whereas using dynamic features, the accuracy is 97\%. 
Our experiment on incorporating both dynamic and static features did not improve the performance of fault detection. Rather it increased the computational overhead (e.g., processing time). Therefore, we excluded static features from the fault detection and categorization phase. However, we found static features to be crucial for fault diagnosis, particularly in conducting root cause analysis. Therefore, we incorporated these features into the explainer module (Section IV-H3).

\setlength{\tabcolsep}{5pt} 
\renewcommand{\arraystretch}{0.92} 
\begin{table}[ht]
\vspace{-1em}
\caption{Performance Comparison Between DEFault \& Baselines}
\centering
\scriptsize 
\begin{threeparttable}
\begin{tabular}{l@{\hskip 5pt}*{5}{c@{\hskip 5pt}c}}
\toprule
\multirow{2}{*}{\textbf{Metrics}} & \multicolumn{2}{c}{\textbf{UM}} & \multicolumn{2}{c}{\textbf{AT}} & \multicolumn{2}{c}{\textbf{DLC}} & \multicolumn{2}{c}{\textbf{DFD}} & \multicolumn{2}{c}{\textbf{DEF}} \\ 
\cmidrule(lr){2-3} \cmidrule(lr){4-5} \cmidrule(lr){6-7} \cmidrule(lr){8-9} \cmidrule(lr){10-11}
 & \textbf{FD} & \textbf{FC} & \textbf{FD} & \textbf{FC} & \textbf{FD} & \textbf{FC} & \textbf{FD} & \textbf{FC} & \textbf{FD} & \textbf{FC} \\
\midrule
\midrule
\multicolumn{11}{c}{\textbf{Test Set}} \\
\midrule
Accuracy   & 0.96 & 0.52 & 0.56 & 0.35 & 0.75 & 0.21 & 0.94 & 0.76 & 0.97 & 0.92 \\
Precision  & 0.95 & 0.54 & 0.60 & 0.37 & 0.74 & 0.23 & 0.94 & 0.75 & 0.97 & 0.92 \\
Recall     & 0.94 & 0.51 & 0.54 & 0.34 & 0.72 & 0.20 & 0.92 & 0.77 & 0.98 & 0.92 \\
F1-Score   & 0.95 & 0.52 & 0.57 & 0.35 & 0.73 & 0.21 & 0.93 & 0.76 & 0.97 & 0.92 \\
\midrule
\multicolumn{11}{c}{\textbf{Benchmark}} \\
\midrule
Accuracy   & 0.91 & 0.27 & 0.23 & 0.19 & 0.69 & 0.13 & 0.90 & 0.52 & 0.94 & 0.63 \\
Precision  & 0.89 & 0.28 & 0.25 & 0.20 & 0.69 & 0.15 & 0.90 & 0.52 & 0.93 & 0.63 \\
Recall     & 0.90 & 0.26 & 0.22 & 0.18 & 0.67 & 0.12 & 0.88 & 0.53 & 0.94 & 0.63 \\
F1-Score   & 0.89 & 0.27 & 0.23 & 0.19 & 0.68 & 0.13 & 0.89 & 0.52 & 0.93 & 0.63 \\
\bottomrule
\end{tabular}
\scriptsize{
\textbf{FD}: Fault Detection, \textbf{FC}: Fault Categorization, \textbf{UM}: UMLAUT, \textbf{AT}: AutoTrainer, \textbf{DLC}: DeepLocalize, \textbf{DFD}: DeepFD, \textbf{DEF}: DEFault
}
\end{threeparttable}
\label{tab:compare}
\vspace{-1em}
\end{table}

\subsubsection{\textbf{Answering RQ$\mathbf{_3}$} -- Comparison with existing baseline techniques} 

We compare DEFault in fault detection and categorization with four established baselines: UMLAUT~\cite{umlaut}, AutoTrainer~\cite{autotrainer}, DeepLocalize~\cite{deeplocalize}, and DeepFD~\cite{deepfd}. Table~\ref{tab:compare} shows that DEFault demonstrates better performance compared to the existing state-of-the-art techniques in both fault detection and categorization tasks. On the testset, DEFault achieves a fault detection accuracy of 97.00\% and a fault categorization accuracy of 92.20\%, outperforming all other techniques. The closest competitor, DeepFD, attains 94.00\% detection accuracy and 76.00\% categorization accuracy. The performance gap is even more significant in real-world evaluation (benchmark), where DEFault has a 94.30\% detection accuracy and 63.46\% diagnosis accuracy. The other baseline techniques exhibit varying degrees of effectiveness. UMLAUT achieves high detection accuracy but struggles with fault diagnosis. However, AutoTrainer and DeepLocalize perform poorly in both detection and diagnosis tasks across both datasets. The limitations of these baseline techniques can be attributed to their approach and scope. UMLAUT, being a heuristic-based framework with specific faults (e.g., learning rate out of common range), may not generalize well to diverse fault scenarios. AutoTrainer focuses on a limited set of training problems, which may not cover all possible fault types. DeepLocalize is limited only to numerical error. In contrast, DEFault demonstrates superior generalization and adaptability by adopting a hierarchical classification framework that learns from a diverse dataset covering all training and model-related faults and leverages dynamic features to detect and categorize faults.

\setlength{\tabcolsep}{4pt} 
\renewcommand{\arraystretch}{1.3}
\begin{table}[ht]
\centering
\vspace{-1em}
\caption{Root Cause Analysis of DEFault \& Baseline Techniques}
\vspace{-1em}
\label{tab:RCA_Layer}
\resizebox{\columnwidth}{!}{
\begin{threeparttable}
\begin{tabular}{p{0.2cm}p{1.45cm}ccc|ccc|ccc|ccc|ccc}
\toprule
\multirow{2}{*}{\#} & \multirow{2}{*}{Fault} & \multicolumn{3}{c}{\textbf{UM}} & \multicolumn{3}{c}{\textbf{AT}} & \multicolumn{3}{c}{\textbf{DLC}} & \multicolumn{3}{c}{\textbf{DFD}} & \multicolumn{3}{c}{\textbf{DEF}} \\
\cmidrule(lr){3-5} \cmidrule(lr){6-8} \cmidrule(lr){9-11} \cmidrule(lr){12-14} \cmidrule(lr){15-17}
 &  & \textbf{FD} & \textbf{FC} & \textbf{RCA} & \textbf{FD} & \textbf{FC} & \textbf{RCA} & \textbf{FD} & \textbf{FC} & \textbf{RCA} & \textbf{FD} & \textbf{FC} & \textbf{RCA} & \textbf{FD} & \textbf{FC} & \textbf{RCA} \\
\midrule
\midrule
1 & act, lay & \cmark & \xmark & \xmark & \cmark & \cmark & \xmark & \cmark & \cmark & \xmark & \cmark & \cmark & \xmark & \cmark & \cmark & \cmark \\
2 & loss, lay & \xmark & \xmark & \xmark & \cmark & \xmark & \xmark & \xmark & \xmark & \xmark & \xmark & \xmark & \xmark & \cmark & \cmark & \xmark \\
3 & act, lay & \cmark & \xmark & \xmark & \xmark & \xmark & \xmark & \xmark & \xmark & \xmark & \cmark & \xmark & \xmark & \cmark & \cmark & \cmark \\
4 & ep, lay & \cmark & \cmark & \xmark & \cmark & \cmark & \xmark & \xmark & \xmark & \xmark & \cmark & \cmark & \cmark & \cmark & \cmark & \cmark \\
5 & lay, act, loss & \cmark & \cmark & \cmark & \xmark & \xmark & \xmark & \cmark & \cmark & \xmark & \cmark & \cmark & \xmark & \cmark & \cmark & \cmark \\
6 & opt, loss, lay & \cmark & \xmark & \xmark & \cmark & \xmark & \xmark & \cmark & \xmark & \xmark & \cmark & \cmark & \xmark & \cmark & \cmark & \cmark \\
7 & ep, lr, act, loss, lay & \cmark & \cmark & \xmark & \cmark & \cmark & \xmark & \xmark & \xmark & \xmark & \cmark & \xmark & \xmark & \cmark & \cmark & \cmark \\
\bottomrule
\end{tabular}
\vspace{.5em}
\end{threeparttable}
}
\scriptsize{
\textbf{UM}: UMLAUT, \textbf{AT}: AutoTrainer, \textbf{DLC}: DeepLocalize, \textbf{DFD}: DeepFD, \textbf{DEF}: DEFault; 
\textbf{FD}: Fault Detection, \textbf{FC}: Fault Categorization, \textbf{RCA}: Root Cause Analysis, 
\textbf{lr}: learning rate, \textbf{act}: activation, \textbf{opt}: optimization, \textbf{ep}: epoch, \textbf{bs}: batch size, \textbf{lay}: layer
}
\vspace{-1em}
\end{table}

\subsubsection{\textbf{Answering RQ$\mathbf{_4}$} -- Assessment of Root Cause Analysis}
To evaluate DEFault's ability to identify the root causes of hyperparameter-related faults (e.g., incorrect batch size), we measured its performance using accuracy, precision, recall, and F1-score. Table \ref{table:performancethirdlevel} shows that DEFault performs well in identifying the root causes of hyperparameter-related faults. With accuracy, precision, recall, and F1-scores consistently above 0.89 for all four hyperparameter fault types (batch size, learning rate, disable batching, and epoch), DEFault demonstrates its effectiveness in pinpointing the specific root causes responsible for the faults. 

Secondly, in our explainer module (Fig. \ref{fig:schematic_diagram}), we focused on the root causes of layer-related faults and utilized the SHAP framework for local explanation~\cite{ghalebikesabi2021locality}. From our benchmark dataset of 52 real-life faulty DNN programs, we collected a total of 7 layer-related faults for our assessment of DEFault. For each of the 7 layer-related faults, we used SHAP to identify the most important features contributing to the detection of faulty behavior. SHAP assigned importance scores to each static feature, allowing us to rank them based on their relevance to the fault. We then compared the top-ranked features identified by SHAP with the ground truth. To assess the performance of our explainer module, we calculated the Top@1 and Top@5 accuracy. These metrics measure the percentage of cases where the correct root cause feature is identified within the top 1 and 5 features ranked by SHAP, respectively. We observed that our explainer module achieved a Top@1 accuracy of 57.10\% and a Top@5 accuracy of 85.71\%. It means for 6 out of the 7 layer faults, the correct root cause feature was present within the top 5 features ranked by SHAP (see Table \ref{tab:RCA_Layer}). SHAP's instance-level explanations enable DEFault to effectively identify the most influential features contributing to layer-related faults, making it suitable for our benchmark dataset, which contains a limited number of such faults. 

\section{Case Study: Evaluation of DEFault}
To demonstrate DEFault's applicability to real-world, complex DNN programs, we conducted a case study using the PixelCNN model from the \textit{Sarus repository} at GitHub~\cite{sarus2023tf2models}. PixelCNN's intricate architecture, resembling real-world applications, makes it a suitable candidate for our case study. We evaluated DEFault using this model on the MNIST~\cite{lecun1998mnist}, a widely used dataset in DL~\cite{kadam2020cnn}, and report our results below.

\subsection{Description of Fault}
The PixelCNN program contained three types of faults in its version \texttt{a83c151}~\cite{sarus2023tf2models}.\\
\textbf{(a) Loss Function Fault:} PixelCNN incorrectly used the \texttt{SparseCategoricalCrossentropy} loss function, which did not align with its objective.\\
\textbf{(b) Hyperparameter Fault:} PixelCNN used an insufficient number of training epochs (i.e., \textit{n}=10 instead of the optimal \textit{n}=75), which adversely affected the model's convergence and learning progress.\\
\textbf{(c) Layer Faults:} The model had misconfigurations in the hidden layer dimensions and residual block counts, which negatively affected its architecture and layers. In particular, the model suffered from these two layer-related issues as follows.
    \begin{enumerate}[wide, labelwidth=!, labelindent=0em,label=(\roman*)]
        \item \textit{Root Cause 1 -- Neuron Count:} The \texttt{hidden\_dim} parameter was set incorrectly (i.e., \textit{n}~=~32 instead of 64). A reduced number of neurons in the layers restricted the model's capacity to capture intricate features within the data.
        \item \textit{Root Cause 2 -- Layer Number:} The \texttt{n\_res} parameter, the number of residual blocks in the model, was incorrect (i.e., \textit{n}~=~3 instead of 6). Fewer residual blocks decreased the network's depth and representational power.
    \end{enumerate}

\begin{figure}[h]
    \centering
        \label{fig:training}
    \vspace{-1em}
    \includegraphics[width=\linewidth]{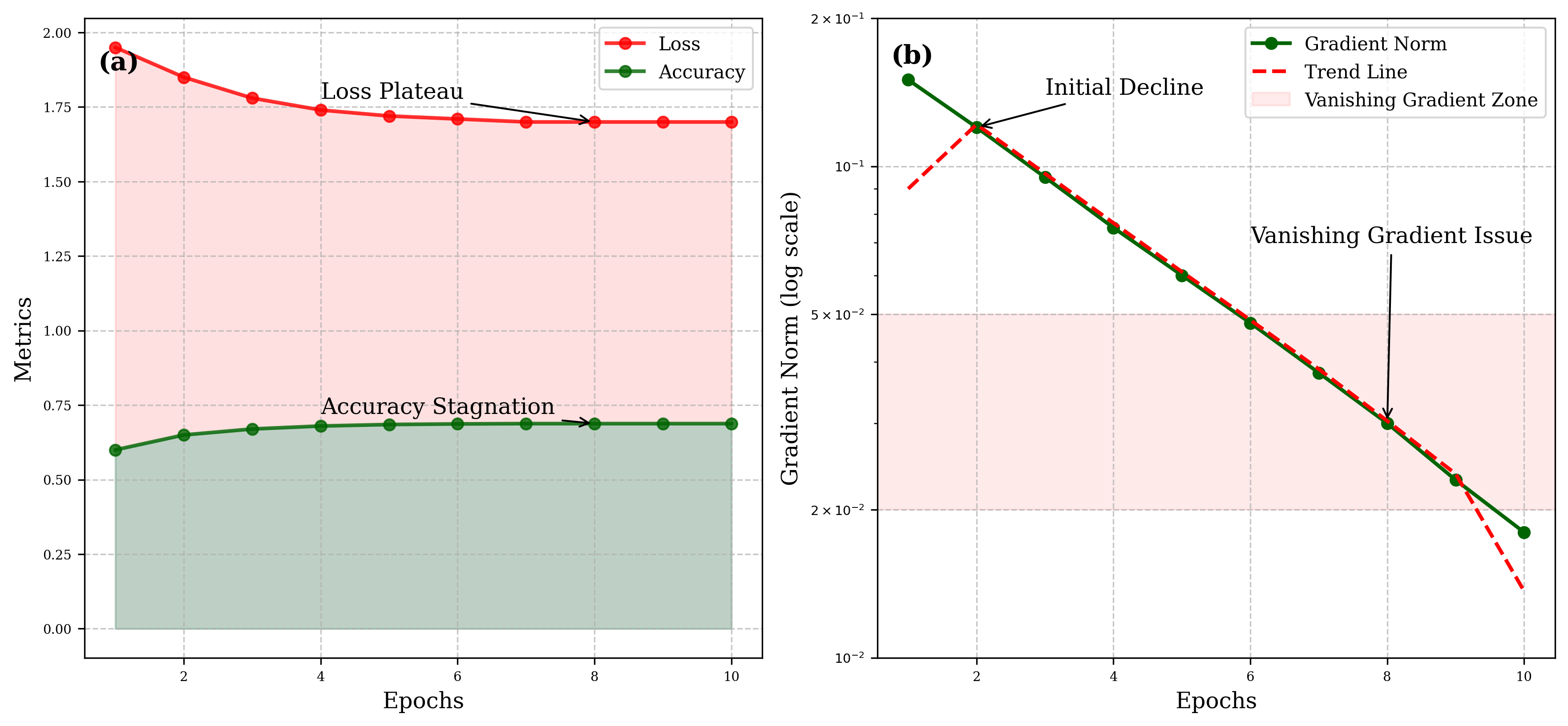}
    \vspace{-2em}
    \caption{(a) Loss-Accuracy (b) Gradient Norms during PixelCNN training}
    \vspace{-1.5em}
\end{figure}

\subsection{DEFault’s Effectiveness \& Result Discussions}

Our proposed technique, DEFault, leverages both static and dynamic information and detects and diagnoses faults in the PixelCNN model in three steps as follows.

\paragraph{{Step 1: Fault Detection}}
DEFault captures dynamic information from the PixelCNN model during its training using our custom callback functions. It monitors runtime features such as loss trends, gradients, and activation statistics. Based on anomalies detected in these metrics (e.g., stagnant loss trends, dead neurons, slow convergence, vanishing gradient), DEFault identified that the DNN program/model (i.e., PixelCNN) contained one or more faults.

\paragraph{{Step 2: Fault Categorization}}
Upon marking the PixelCNN model as faulty, DEFault attempts to categorize its faults using seven binary classifiers. It leverages dynamic features collected from our custom callbacks during training. According to our experiments, DEFault successfully identified the presence of three faults: Loss Function Fault, Hyperparameter Fault, and Layer Fault in the PixelCNN model.

\paragraph{{Step 3: Root Cause Analysis}}
DEFault also analyzes the root causes of the detected faults, leveraging dynamic and static features as follows. \\
\indent
\textit{Step 3.1: Hyperparameter Fault:} DEFault captures several dynamic features, such as loss and accuracy trends, from the PixelCNN model during its training and analyzes their importance or relevance to the hyperparameter faults. Using the Level 3 classifier, our technique also marks the \textit{number of epochs} as the root cause for the hyperparameter faults. Our post-hoc analysis (Fig. 6a) suggests that the training loss decreased slightly during the initial epochs but then stagnated. DEFault's continuous monitoring of loss values within an epoch and across multiple epochs might have helped capture the stagnated loss trends and identify the loss function fault (Level 2 classifier, Step 2). We also notice the stagnant accuracy throughout the training process of the PixelCNN model, suggesting incomplete model convergence. By capturing these training behaviors (e.g., stagnant loss and accuracy) at regular intervals, Default's Level 3 classifier confirms the presence of an epoch fault. Given the lack of model convergence, the number of epochs might not be sufficient, which explains DEFault's root cause behind the hyperparameter faults.\\
\indent
\textit{Step 3.2: Layer Fault:} DEFault captures several dynamic features, such as activation and gradients, from the PixelCNN model during its training and analyzes their importance and relevance to the layer faults. DEFault captures static features (e.g., layer numbers, neuron count) from the PixelCNN model to diagnose the layer faults detected above. It conducts a SHAP-based analysis of the static features leveraging its Explainer Module (see Section IV-G) and generates the following explanations of the root causes. \\
\indent \textbf{Top@1 CountDense:} Check the configuration and number of Dense layers. \\
\indent \textbf{Top@2 Max\_Neurons:} Verify the maximum number of neurons in any single layer. \\
\indent \textbf{Top@3 CountConv2D:} Inspect the configuration of 2D convolutional layers.

All these messages suggest the number of hidden layers and the number of neurons as the potential root cause of the layer fault. Our post-hoc analysis suggests that during the training of the PixelCNN model, many neurons remained inactive with outputs at zero, indicating a higher percentage of `dead neurons'. We also observed abnormally low gradient norms (see Fig. 6b), suggesting vanishing gradients in the model. DEFault's continuous monitoring of activation and gradient statistics within an epoch and across multiple epochs might have helped it capture activation saturation or vanishing gradients and identify the layer faults (Level 2 classifier, Step 2). We also noticed that static features such as \texttt{CountDense}, \texttt{Max\_Neurons}, and \texttt{CountConv2D} are the most important in the explainer module targeting the layer faults. This also indicates that misconfigurations in the layers and neurons might have triggered the bug, which explains DEFault's identification of these aspects as the root cause behind the layer faults.

\subsection{Limitations of DEFault}
Although DEFault correctly identified three types of faults in the PixelCNN model using its hierarchical classifier, it misclassified an optimization fault. Fig. 6a shows that the faulty loss function in PixelCNN negatively affects the optimization process, resulting in stagnant loss and accuracy trends. Since optimization parameters are linked to loss calculation~\cite{bruch2019analysis}, which is faulty in PixelCNN, DEFault might have predicted the faulty optimization inaccurately. To understand this misclassification, we analyzed the top features influencing DEEault's loss and optimization classifiers. We found that features like \texttt{gpu/cpu\_utilization}, \texttt{train\_acc}, and \texttt{memory\_usage} are influential in both classifiers, suggesting their overlapping behavior. Furthermore, our analysis reveals a significant correlation among their features. For instance, the negative correlation between \texttt{val\_acc} and \texttt{adj\_lr} ($-0.95$) highlights a strong connection between loss/accuracy calculation and the optimization process. Thus, DEFault might have difficulty distinguishing between loss and optimization faults due to their overlapping and strongly connected behaviors.

\vspace{-.5em}
\subsection{Comparison with Baseline Techniques}
To validate DEFault's effectiveness in fault detection and diagnosis, we compared it with four baseline techniques. DeepFD detected the hyperparameter fault and loss function fault but missed the layer fault. This limitation could be attributed to its focus on anomalies in specific dynamic features, neglecting activation-related dynamic features, which our experiments found to be relevant for identifying layer faults. Additionally, its pre-trained model lacks support for layer faults and does not account for architectural properties (e.g., static features), further contributing to this shortcoming. Moreover, DeepFD also misclassified optimization as a fault like DEFault, reinforcing our observation regarding the overlap between loss calculation and optimization. 

UMLAUT relies on heuristics (e.g., validation accuracy trends) to map between symptoms and root causes of DNN faults. Using the heuristics, it was able to detect hyperparameter faults, but it could not pinpoint the root cause of the fault (e.g., number of epochs). It also failed to detect the loss function fault and layer faults, suggesting the limitations of heuristics against structural issues in the DNN model. 

Autotrainer captures dynamic metrics from a DNN model during training and uses the rule-based technique to detect faults and repair the model. While it correctly identified the oscillating loss behavior caused by the incorrect loss function, it could not detect the hyperparameter fault or layer faults. This is likely because its rule-based design only looks for patterns that match the predefined symptoms for specific training problems. Hyperparameter faults, like insufficient epochs and layer faults, may not directly match these predefined patterns, which limits Autotrainer's ability to identify issues that it was not designed for.

DeepLocalize was designed to detect numerical errors like NaN or infinity in loss or activations, which were absent in the PixelCNN faults. Thus, it was not able to detect any of the three bugs from the PixcelCNN model. 

Overall, DEFault demonstrated broader fault coverage by successfully identifying both runtime (e.g., loss function fault, hyperparameter fault) and structural issues (e.g., layer faults) in the PixelCNN model where the contemporary baselines fall short due to their limited scopes or specific focus areas.

\section{Threats To Validity}
Threats of \emph{internal validity} relate to experimental errors. Re-implementation of the existing baseline techniques could pose a threat. Since we used the replication packages provided by the original authors for all our baseline models, such a threat might be mitigated. We also repeat our experiments 15 times and compare the performance with that of baselines to mitigate any bias due to random trials~\cite{arcuri2014hitchhiker}.

Threats to construct validity~\cite{smith2005construct} relate to the subjectivity in manual assessment.
Filtration and reproduction of DNN programs during benchmark construction could pose such a threat. To minimize individual biases, two authors independently examined each subject, and we see a high agreement level between them (e.g., 85.4\% Cohen Kappa score).

Threats to \emph{external validity} relate to the generalizability of our work~\cite{findley2021external}. Whether the dataset represents real-world faults or not is such a threat. Since we collected all DNN-related posts from StackOverflow and filtered them through a rigorous set of criteria following existing literature, the threat is minimal. Whether the selected performance metrics measure the research problem properly or not is another threat. The literature well-defines all of our performance metrics, and the selection is supported by the baseline literature. Thus, the threat to external validity is also minimal.

\section{Related Work}
\textbf{Mutation Testing:} To overcome the scarcity of publicly available faulty DNN programs, prior works~\cite{deepfd} have used synthetic models (i.e., mutant models) generated by DeepCrime~\cite{deepcrime} as the source of the training dataset. However, DeepCrime might be limited since it supports six out of seven major categories from the literature \cite{faulttaxonomy, dlbugcharacterstics}. It also does not support RNN-based mutation. Another mutation generation technique, DeepMutation~\cite{deepmutation}, only supports three root causes in feed-forward neural networks. The limited mutation coverage of these techniques may result in datasets containing only a subset of DNN faults. DEFault addresses this limitation by creating an extended version of DeepCrime, allowing improved mutation coverage and enabling DEFault to be trained on a comprehensive set of possible faults and root causes. 

\textbf{Debugging DNN Programs:} Recent research has proposed various techniques for diagnosing faults in DNN programs. Static analysis-based techniques, such as Neuralink~\cite{Neuralink}, NerdBug~\cite{nerdbug}, and DEBAR~\cite{debar}, analyze code structure and syntax without execution to detect faults, but may miss runtime-specific issues. On the other hand, dynamic analysis-based techniques, including UMLAUT~\cite{umlaut}, AutoTrainer~\cite{autotrainer}, DeepDiagnosis~\cite{deepdiagnosis}, and DeepLocalize~\cite{deeplocalize}, use runtime information from the training session to detect faults. However, they rely on predefined rules that limit their ability to generalize. DeepFD~\cite{deepfd} addresses these limitations by using dynamic features in ML classifiers for fault detection but overlooks structural model-related faults~\cite{dlbugcharacterstics}. DEFault presents a \textit{novel} approach that combines static and dynamic features in a hierarchical classification approach. It addresses the limitations of existing techniques by detecting, categorizing and root cause analyses of both training and model-related faults.

\section{Conclusion}
In this paper, we present DEFault, a novel technique for detecting and diagnosing faults in DNN programs. Our experiments on a diverse dataset of 14,652 DNN programs and real-world faulty models from StackOverflow show that DEFault outperforms state-of-the-art techniques in fault detection and diagnosis of DNN programs. Moreover, the explainer module of DEFault effectively determines the root causes for model-related faults using static features. Our proposed dynamic features further enhance DEFault's performance in fault detection. Future work includes expanding the prototype to handle more architecture types (e.g., attention-based neural networks) and automatically fixing bugs.

\printbibliography
\end{document}